\documentclass[a4paper,11pt]{article}

\pdfoutput=1
\usepackage{booktabs}

\usepackage{lineno}
\usepackage[colorlinks=false]{hyperref}
\usepackage{bm}
\usepackage{graphicx}
\usepackage{amssymb}
\usepackage{amsmath}
\usepackage{tabularx}
\usepackage{natbib}
\usepackage{floatrow}
\usepackage{caption}
\usepackage{multirow}
\usepackage{amsmath}
\usepackage{amsfonts}
\usepackage{amssymb}
\usepackage{xcolor}
\usepackage{subfig}
\usepackage{graphicx,amsmath,bm}
\usepackage[binary-units=true]{siunitx}
\usepackage{array}
\usepackage{authblk}
\usepackage{rotating}
\usepackage{enumerate}
\usepackage{ragged2e}

\usepackage[left=15mm,right=15mm,top=1.5cm,bottom=1.5cm,includeheadfoot]{geometry}
\setcitestyle{square,numbers}
\begin{document}
	
	\title{Quality-factor inspired deep neural network solver for solving inverse scattering problems}
	
	\author[1]{Yutong~Du}
	\author[1]{Zicheng~Liu}
	\author[1]{Miao~Cao}
	\author[1]{Zupeng~Liang}
	\author[1]{Yali~Zong}
	\author[1]{Changyou~Li}
	
	\affil[1]{\scriptsize Department of Electronic Engineering, School of Electronics and Information, Northwestern Polytechnical University, Xi'an 710029, China}
	\maketitle
	
	\abstract{
		Deep neural networks have been applied to address electromagnetic inverse scattering problems (ISPs) and shown superior imaging performances, which can be affected by the training dataset, the network architecture and the applied loss function. Here, the quality of data samples is cared and valued by the defined quality factor. Based on the quality factor, the composition of the training dataset is optimized. The network architecture is integrated with the residual connections and channel attention mechanism to improve feature extraction. A loss function that incorporates data-fitting error, physical-information constraints and the desired feature of the solution is designed and analyzed to suppress the background artifacts and improve the reconstruction accuracy. Various numerical analysis are performed to demonstrate the superiority of the proposed quality-factor inspired deep neural network (QuaDNN) solver and the imaging performance is finally verified by experimental imaging test.}
	
	\section{Introduction}
	%
	%
	%
	%
	Electromagnetic inverse scattering imaging \cite{chen2018computationalEMIS} is a technique for image reconstruction that exploits the complex interactions between electromagnetic wave and matter. Inverse scattering holds significant promise for successful applications, including early tumor detection \cite{dachena2021tumordetection} and cancer diagnosis \cite{hirose2022cancerdiagnosis}, non-destructive testing of defects such as microcracks and porosity in composite materials \cite{An2024Composites}.  Due to multiple scattering effects, nonlinear ISPs need to be solved and may stuck into local optima. The inherent ill-posedness leads to challenges in solver stability when considering measurement uncertainties \cite{chen2018computationalEMIS}. Such challenges lead to the requirement of accurate and reliable reconstruction schemes.
	
	Quantitative analysis \cite{bucci1997quantitativeanalysis,Adrian1988quant,A1988quant,Stefanov1990quant,P2009quant} of ISPs aims to reconstructing the electrical properties of scatters in detection and the related classical algorithms can be categorized into iterative and non-iterative methods. Non-iterative methods, including the Born approximation (BA) method \cite{habashyi1993BA,gao2006BA}, the Rytov approximation (RA) method \cite{slaney1984RA,devaney1981RA,alon1993RA}, and the backpropagation (BP) method \cite{devaney1982BP,tsili1998BP,Belkebir2005BP}, ignore high-order scattering effects and is lacking the ability of reconstructing the high-contrast scatters and the details. Iterative inversion methods, including Born iterative method (BIM) \cite{wang1989BIM,sung1999BIM}, distorted Born iterative method (DBIM) \cite{chew1990DBIM,haddadin1995DBIM}, contrast source inversion (CSI) method \cite{peter1997CSI,richard2001CSI} and subspace optimization method (SOM) \cite{chen2010SOM,chen_2010SOM,pan2011SOM}, exhibit strong robustness and capability of handling high-contrast scatters, but suffer from high computational cost and the resulting low reconstruction efficiency.
	
	Deep learning \cite{li2019DNN,wei2019DNN} has been shown superior in dealing with ISPs by predicting contrast through data-driven convolutional neural network (CNN) \cite{Chen2023CNN,Wu2024CNN}. However, the obtained solvers are regarded as a ``black box” which establish the nonlinear relationship between input and response without explicit constraints from physical laws. As a result, the obtained solvers need improvements in generalization ability. 
	
	Incorporating physical constraints into the learning process is key to improve the generalization ability. Wei et al. take the estimation from BP as input to making use of the physical information \cite{wei2019DNN}. Yao et al. proposed a two-step enhanced deep learning approach for ISPs, where the features of scattered fields are learned and are used to represent the initial image of scatterers with a neural network \cite{yao2019DNN}. Liu et al. proposed physics-guided loss functions that additionally constrain near-field quantities (scattered field and induced current in the domain of interest) to enhance the noise robustness and scatter features \cite{liu2022DNN}. Ma et al. leveraged the smoothness of the contrast distribution by incorporating a total variation (TV) penalty into the loss function \cite{ma2023DNN}. 
	
	To improve the imaging performance further, the following efforts are made in this paper: 
	
	(1) Sample quality is considered to optimize the composition of training dataset. Usually, the training dataset is generated with simulation tools with virtual scatterers, the relative permittivity of which is randomly selected from a specific range. Such generation way is simple but neglects that fact that samples can have different impacts on the trained model. In general, samples that are challenging to predict include more information for the training process. In the paper, the authors defined a quality factor which evaluates the contribution of sample based on the prediction performance by a traditional ISP solver. Based on the quality factor, the composition of training dataset is designed to include more low-quality samples rather than samples that can be easily reconstructed.    
	
	(2) The loss function which combines the physical information and the data fitting one is designed. Despite the contrast discrepancy term, the loss function applied integrates the near-field physical constraints \cite{liu2022DNN} and the smooth character of the desired imaging results \cite{ma2023DNN}. In addition, to evaluate the imaging accuracy fairly, both RMSE and SSIM are used to quantify the Euclidean distance and structural similarity between the predicted and the ground truth, respectively. 
	
	(3) This paper presents a deep neural network (DNN) architecture ReSE-U-Net, which incorporates residual connections \cite{He2016residual}, channel attention mechanism \cite{Hu2018SE} and the so-called feature transformation layer into the traditional U-Net architecture \cite{ronneberger2015UNet}. The design aims to alleviate the problem of gradient vanishing and explosion in DNNs, and leading to performance improvements in deep architectures.
	
	The paper is organized as follows. Section \ref{sec:formulateEMIS} introduces the concerned inverse source problems. In Section \ref{sec:DNN solvers}, the definition of quality factor is introduced and used to decide the composition of training dataset. Besides, the DNN architecture ReSE-U-Net and the designed loss functions are described. In Section \ref{sec:analyze data}, numerical analysis are made to compare the imaging performances with different training settings and show the superiority of the proposed training scheme. Conclusions are made in Section \ref{sec:conclusions}.
	
	The mathematical notations are used by denoting variables in italic font, putting bars on vector and matrix notations, \emph{e.g.}, $\mathbf{\bar{X}}$ and $\mathbf{\bar{\bar{X}}}$. The Euclidean length is denoted by $\Vert\cdot\Vert$, and the diagonalization operator is represented as diag($\cdot$). $\mathbf{I}$ is the identity matrix.
	\begin{figure}
		\centering
		\includegraphics[width = 0.4\linewidth]{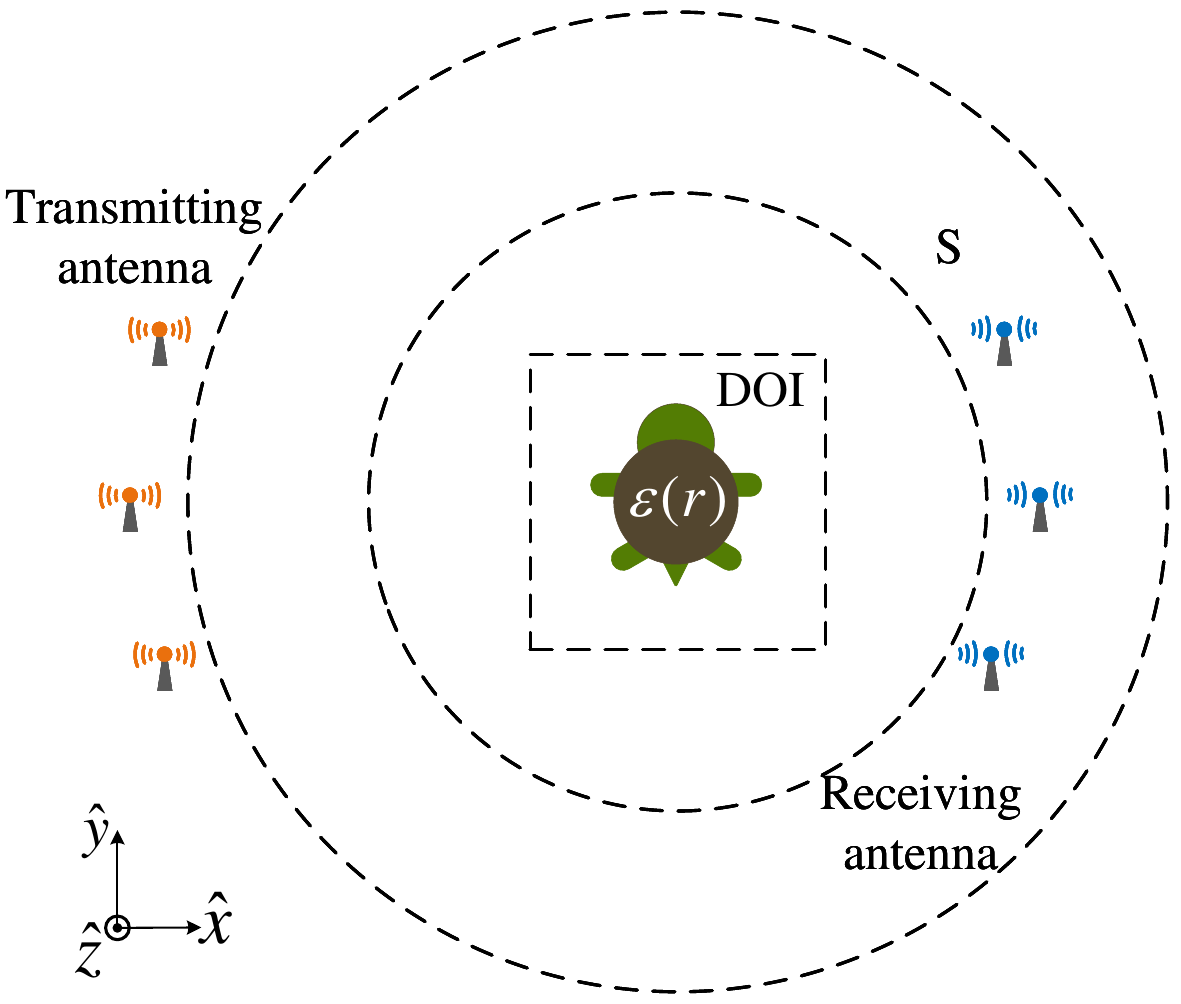}
		\caption{The imaging scheme of the concerned two-dimensional inverse scattering problem.}
		\label{fig:sketchISP}
	\end{figure}
	
	\section{Formulation of ISPs}
	\label{sec:formulateEMIS}
	The concerned imaging scheme is shown in Fig.~\ref{fig:sketchISP}, where a two-dimensional transverse-magnetic (TM) wave illumination with polarization along the $z$-axis is considered \cite{meaney1995TM}. The background is homogeneous with relative permittivity ${\epsilon_0}$ and permeability ${\mu_0}$. Non-magnetic scatterers are located within domain of interests (DOI). Scatter information is collected with $\mathrm{N_{i}}$ line sources, which are located at ${\bf{r}}_p^i$, $p=1,2,...,\mathrm{N_{i}}$, and $\mathrm{N_{s}}$ receiving antennas located at a surface S. 
	
	The forward problem can be described by data equation and state equation. The state equation characterizes the interactions between the incident field and the scatterers \cite{chen2018computationalEMIS}
	\begin{equation}
		\mathbf{E}^\text{tot}_\text{DOI}(\mathbf{r}) = \mathbf{E}^\text{inc}_\text{DOI}(\mathbf{r}) + k_0^2\int_\text{DOI}g(\mathbf{r},\mathbf{r}^\prime)\mathbf{J}(\mathbf{r}^\prime)d\mathbf{r}^\prime,\mathbf{r}\in\text{DOI},
		\label{eq:stateEqu}
	\end{equation}
	where $\mathbf{E}^\text{tot}_\text{DOI}(\mathbf{r})$ and $\mathbf{E}^\text{inc}_\text{DOI}(\mathbf{r})$ denote the total and incident electric field at the observation point $\mathbf{r}$, respectively. $k_0$ stands for the wavenumber of the background medium. $g$ is the two-dimensional scalar Green's function. The induced current $\mathbf{J}$ is defined as 
	\begin{equation}
		\mathbf{J}(\mathbf{r}^\prime) = \boldsymbol{\chi}(\mathbf{r}^\prime)\mathbf{E}^\text{tot}(\mathbf{r}^\prime),
		\label{eq:nonlinearEqu}
	\end{equation}
	where $\boldsymbol{\chi}(\mathbf{r}^\prime) = \epsilon_r(\mathbf{r}^\prime)-1$ and $\epsilon_r(\mathbf{r}^\prime)$ denotes the relative permittivity at the observation point $\mathbf{r}^\prime$. The induced current radiates fields that are measured by receiving antennas which can be quantified by data equation, \emph{i.e.},
	\begin{equation}
		\mathbf{E}^\text{sca}_\text{mea}(\mathbf{r}) = k_0^2\int_\text{DOI}g(\mathbf{r},\mathbf{r}^\prime)\mathbf{J}(\mathbf{r}^\prime)d\mathbf{r}^\prime, \mathbf{r}\in\text{S}.
		\label{eq:dataEqu}
	\end{equation} 
	Equations (1) and (3) can be rewritten as
	\begin{equation}
		\mathbf{E}^\text{tot}(\mathbf{r}) = \mathbf{E}^\text{inc}(\mathbf{r}) + \mathbf{G}_\text{D}(\boldsymbol{\chi}\mathbf{E}^\text{tot}), \,\,\mathbf{for} \,\,\mathbf{r}\in\text{DOI},
		\label{eq:GreenOperatorStateEqu}
	\end{equation}
	\begin{equation}
		\mathbf{E}^\text{sca}(\mathbf{r}) = \mathbf{G}_\text{S}(\boldsymbol{\chi}\mathbf{E}^\text{tot}), \,\,\mathbf{for} \,\,\mathbf{r}\in\text{S},
		\label{eq:GreenOperatorDataEqu}
	\end{equation} 
	
	Method of moments (MoM) \cite{Ney1985MoM,Gibson2021MoM,LAKHTAKIA1992MoM} with the impulse basis function and the delta test function is applied to solve (4) and the scattered fields in (5) can be described as
	\begin{equation}
		\mathbf{E}^\text{sca}(\mathbf{r}) = \mathbf{G}_\text{S}\boldsymbol{\chi}(\mathbf{I}-\mathbf{G}_\text{D}\boldsymbol{\chi})^\mathbf{-1}\mathbf{E}^\text{inc}.
		\label{eq:FinalDataEqu}
	\end{equation} 
	
	For inverse problems, regularization method \cite{golub1999Tikhonov, Leonid1992TV} is used to increase the stability of solution which is obtained by solving the following optimization problem 
	\begin{equation}
		\min: L({\bar{\boldsymbol{\chi}}}) = \sum_{p=1}^\mathbf{N_i}||\mathbf{E}^\text{sca}_{p}-\bar{\mathbf{E}}^\text{sca}_{p}({\bar{\boldsymbol{\chi}}})||^2 + \alpha Q({\bar{\boldsymbol{\chi}}}),
		\label{eq:optimizationEq}
	\end{equation}
	where $\bar{\mathbf{E}}^\text{sca}_{p}({\bar{\boldsymbol{\chi}}})$ denote scattered fields w.r.t. the contrast distribution $\bar{\boldsymbol{\chi}}$ and ${\mathbf{E}}^\text{sca}_{p}$ is the measured ones. $Q({\bar{\boldsymbol{\chi}}})$ is the regularization term. ${\alpha}$ is the parameter trading-off the contributions from the data-fitting term and the regularization one. 
	
	\section{Deep neural network solvers}
	\label{sec:DNN solvers}
	
	Here, one concerns on the quality of samples and trys to optimize the composition of training dataset so that more informative samples are included. To improve the feature learning ability, residual connections, Squeeze-and-Excitation (SE) blocks and feature transformation layer are integrated in the traditional U-Net architecture. The loss function is also designed that both the data-fitting error, physical-information constraint and the desired solution smoothness are considered together.
	
	\begin{figure}[!t]
		\centering
		\includegraphics[width = 0.5\linewidth]{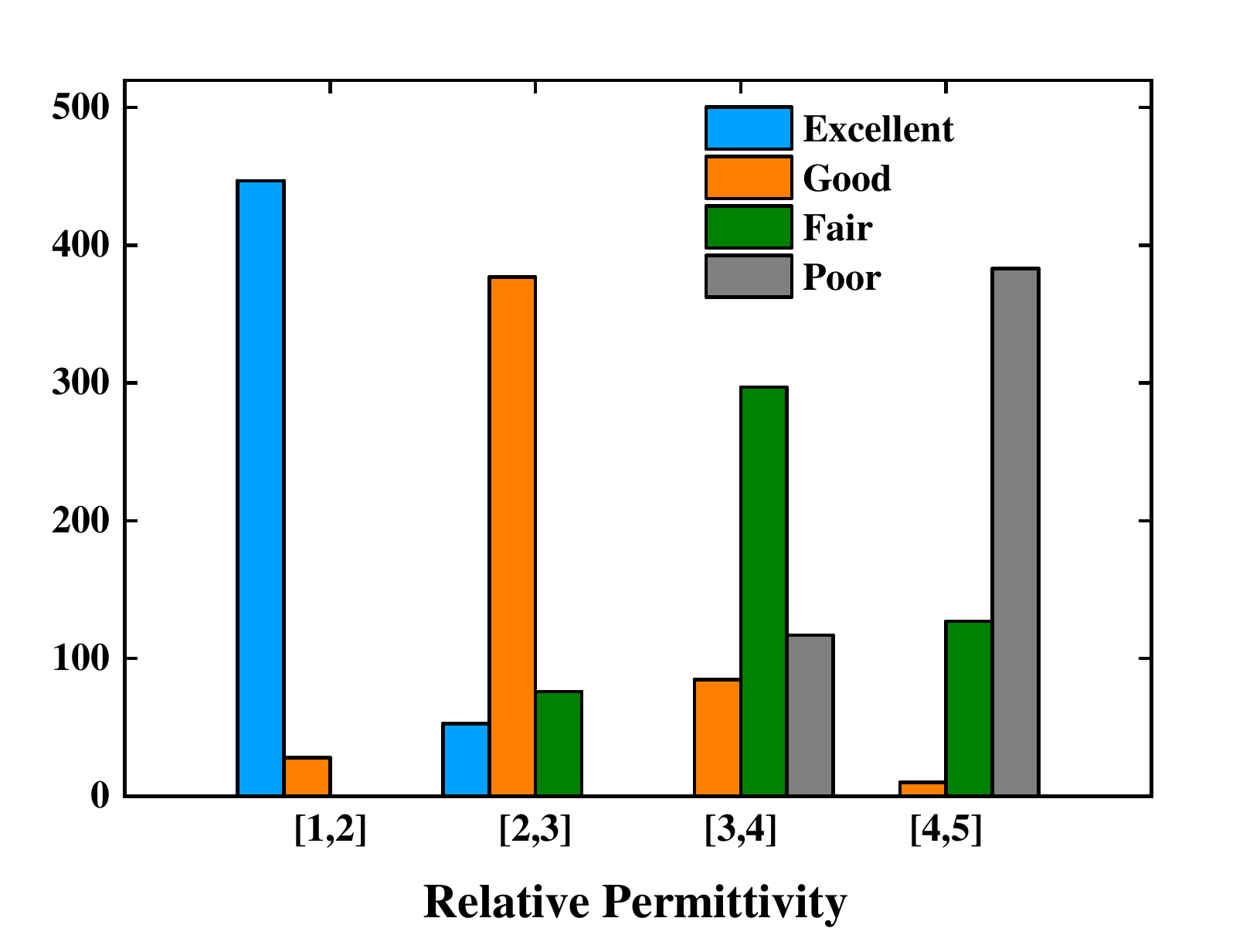}
		\caption{Distribution of samples according to quality factor and relative permittivity value.}
		\label{fig:datadistribution}
	\end{figure}
	
	\begin{figure}[!t]
		\centering
		\includegraphics[width = 0.6\linewidth]{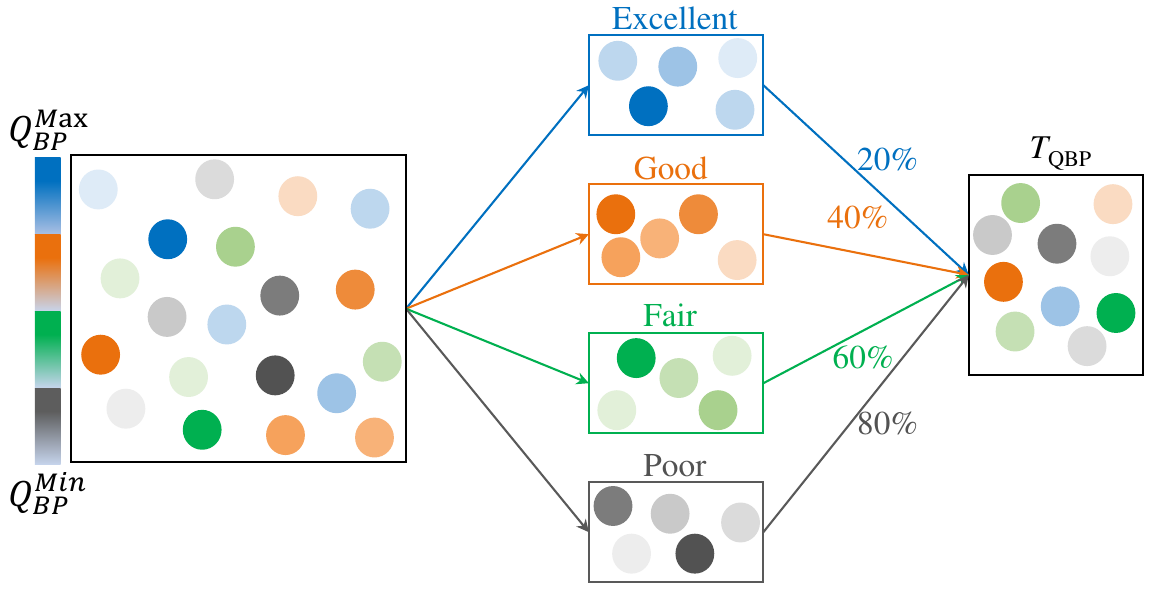}
		\caption{Composition of training dataset $T_{\text{QBP}}$ according to quality factor $Q_{\text{BP}}$.}
		\label{fig:trainingdataset}
	\end{figure}
	
	\subsection{Quality factor and training dataset composition}
	\label{subsec:dataGenerationandQualityFactor}
	The training dataset is constructed based on MNIST database \cite{Lecun1998MNIST}, which consists handwritten digit images. The DOI is with size $5.6\lambda_0 \times 5.6\lambda_0$, where the working wavelength $\lambda_0 = \SI{7.5}{\centi\meter}$, and discretized into $64 \times 64$ grids. With linear transformation, the digit images are transformed as scatterers with relative permittivity in the range of $(1,5]$. Since low SNR ISPs are focused here and noisy training data are helpful in noise robustness, white Gaussian noise is added to the obtained scattered fields. For real-time imaging, non-iterative BP method is used to generate the initial image of the target scatterers, which is the input of the neural network.
	
	The generation method of training dataset is optimized based on the defined quality factor $Q_{\text{BP}}$
	\begin{equation}
		Q_{\text{BP}} = \frac{\text{SSIM}}{\text{RMSE}}
		\label{eq:QBP}
	\end{equation}
	where SSIM is short for structure similarity index measure and RMSE for root mean square error. The subscript ``BP" denotes that SSIM and RMSE are estimated based on the ground truth and the estimated image from back propagation method. Since higher SSIM and lower RMSE indicate more accurate prediction, larger $Q_{\text{BP}}$ implies higher-quality data sample. 
	
	The reasonability of using $Q_{\text{BP}}$ to quantify the sample quality is analyzed. Randomly choosing 2000 MNIST samples, and using $36$ transmitters and receivers to collect the scattered fields, the $Q_{\text{BP}}$ w.r.t. each sample is calculated after having the BP imaging results. Resorting the samples in descending order according to $Q_{\text{BP}}$, the quality of samples are categorized as ``excellent", ``good", ``fair", and ``poor", each ground composed of 500 samples. BP tends to underestimate the permittivity values\cite{liu2022DNN}. Such phenomena are checked with the distribution of $Q_{\text{BP}}$. Fig.~\ref{fig:datadistribution} shows the number of samples with different quality labels within the contrast interval (1,2], (2,3], (3,4], (4,5], respectively. In the (1,2] interval, the majority of samples are classified as ``excellent", while the majority of samples fall into the `good" category for the (2,3] interval, `fair" category for the (3,4] interval, and ``poor" category in the (4,5] interval. The observations suggest the reasonability of $Q_{\text{BP}}$ definition and also revealed a close relationship between the quality of samples and the range of relative permittivity. 
	
	Here, the composition of training dataset is optimized according to $Q_{\text{BP}}$. The training dataset is composed of different-quality samples, but the majority is from the ``poor" category considering that the poorly estimated samples by BP should be more effectively learned with more samples. The proportion of the four categories, \emph{i.e.}, 10\% ``excellent" set, 20\% ``good" set, 30\% ``fair" set and ``40\%" poor set, is found effective for the dataset composition. With the proportion limit, as skectched in Fig.~\ref{fig:trainingdataset}, randomly choose 1000 samples to compose the dataset ``$T_{\text{QBP}}$"  and denote the randomly selected dataset from a uniform distribution as ``$T_{\text{UNI}}$". 
	
	\subsection{ReSE-U-Net Architecture}
	\label{subsec:ReSE-U-Net}
	
	\begin{figure*}[!th]
		\includegraphics[width=\linewidth]{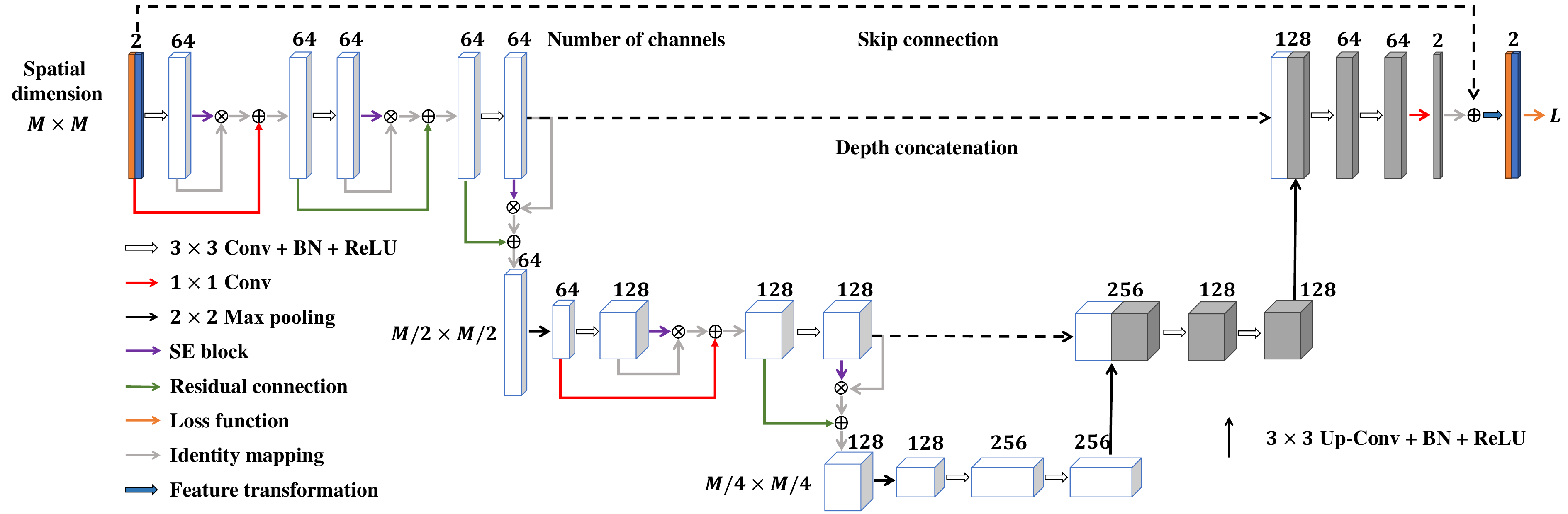}
		\caption{ReSE-U-Net architecture used to solve ISPs in this paper. The input consists of two channels representing the real and imaginary parts of BP imaging result.The architecture integrates residual connection, channel attention mechanism and feature transformation layer into U-Net.}
		\label{fig:DNNArchitecture}
	\end{figure*}
	
	The ReSE-U-Net architecture is an modified U-Net \cite{ronneberger2015UNet} and shown in Fig.~\ref{fig:DNNArchitecture}. The neural network consists of input layer, encoder, decoder and output layer. The real part and the imaginary part of complex-valued input are treated as two independent channels. In the encoder section, each convolutional block contains a $3\times3$ convolutional layer, batch normalization (BN), and ReLU activation function, followed by the Squeeze-and-Excitation (SE) block \cite{Hu2018SE} and residual connections \cite{He2016residual}. The convolutional blocks extract low-level features from the input image, and the SE blocks impose channel attention mechanism that adaptively recalibrates channel-wise feature responses by explicitly modeling inter-dependencies between channels. By incorporating $2\times2$ max pooling, the feature map dimension is halved while preserving key information. In the decoder section, each transposed block contains a $3\times3$ up-convolution layer, BN, and ReLU activation function. A feature transformation layer consisting of $3\times3$ convolutional layer and BN is introduced to reduce numerical instability, allowing the model to have more stable training results in complex tasks. The skip connection helps the decoder retain spatial detail information from earlier stages of the encoder, facilitating more accurate feature recovery. 
	
	The inclusion of SE blocks, residuals and feature transformation layer is beneficial in solving ISPs for the following reasons:
	\begin {enumerate}
	\item Preventing degradation problems in DNNs. In our setup for dealing ISPs, both the inputs and outputs of the DNN represent relative permittivity, ensuring consistent physical scale. The training dataset includes input samples, some of which are close to the ground truth and some having significant deviations. For the input samples close to the ground truth, over-training may lead to the increase of training error. With the residual connections, if a layer is unable to learn valid information, the network can still propagate the input through the skip connections, thereby preventing performance degradation. 
	\item Avoiding gradient vanishing in DNNs. During the backpropagation of the neural network, gradients may either diminish or explode as the number of layers increases, leading to excessively slow or unstable weight updating. Residual connections can mitigate the vanishing gradient problem by allowing gradients to bypass intermediate layers and propagate directly, facilitating more stable and efficient training.
	\item Improving the noise robustness. Under low SNR conditions, the SE block can adaptively recalibrate channel-wise feature responses by explicitly modeling inter-dependencies between channels. This mechanism enables the network to prioritize the most relevant input features while suppressing irrelevant or noise-affected ones. This selective enhancement mechanism indirectly improves the ability of the network to handle noisy data, leading to performance improvements in deep architectures with minimal additional computational cost.
	\end {enumerate}
	
	\subsection{Loss Functions}
	\label{subsec:loss functions}
	The loss function quantifying the Euclidean distance between the predicted contrast and the ground truth, \emph{i.e.},
	\begin{equation}
		L^{\text{contrast}} = ||\hat{\boldsymbol{\chi}}-\boldsymbol{\chi}||^2
		\label{lossL2norm}
	\end{equation}
	is commonly utilized to optimize the network weights. $\boldsymbol{\chi}$ and $\hat{\boldsymbol{\chi}}$ denotes the true contrast distribution and the predicted one, respectively. To combine the visual imaging quality criterion, the following loss function is proposed  
	\begin{equation}
		L^{\text{SSIM}} = L^{\text{contrast}} + (1-{\text{SSIM}}^2)
		\label{lossLQBP}
	\end{equation}
	
	To integrate the physical constraint and make use of the smoothness of the desired image, the loss function 
	\begin{equation}
		L^{\text{mix}} = L^{\text{SSIM}} + \alpha L^{\text{field}} + \beta L^{\text{TV}}
		\label{lossLmix}
	\end{equation}
	is designed, where $L^{\text{field}}$ comes from \cite{liu2022DNN}, and $L^{\text{TV}}$ is the TV regularization term. $\alpha$ and $\beta$ are the hyperparameters, where $\alpha = {||\boldsymbol{\chi}||^2}/{||\mathbf{E}^\text{sca}_\text{DOI}||^2}$ to balance the contributions, and $\beta$ is within the range $[0, 1]$ .
	
	\section{Numerical Result}
	\label{sec:analyze data}
	\subsection{Training Parameter Settings}
	\label{subsec:DNNset}
	The models are trained with a workstation equipped with 128 GB RAM, 3.20 GHz CPU, and RTX 4090 GPU. The stochastic gradient descent algorithm with momentum 0.99 is applied. The initial learning rate is set to $5\times 10^{-6}$, and the training process terminates when the number of epochs reaches 150. 
	
	The scatterers under detection are with relative permittivity ranged from 1 to 5. 1000 digit-like profiles are randomly selected as the training dataset $T_{\text{UNI}}$, and the training dataset $T_{\text{QBP}}$ is generated according to the quality factor $Q_{\text{BP}}$. To test the generalization ability, an independent testing dataset composed of 2000 polygon-like scatterers was generated by randomly positioning regular polygons with 3 to 7 sides (allowing overlaps) within the DOI. The distance from the vertices to the geometric center was randomly selected between 0.1$\lambda_0$ and 1.6$\lambda_0$. 
	
	\subsection{U-Net vs. ReSE-U-Net}
	\label{subsec:TestReSE-U-Net}
	\begin{figure*}[!t]
		\begin{center}
			\includegraphics[width=0.87\linewidth]{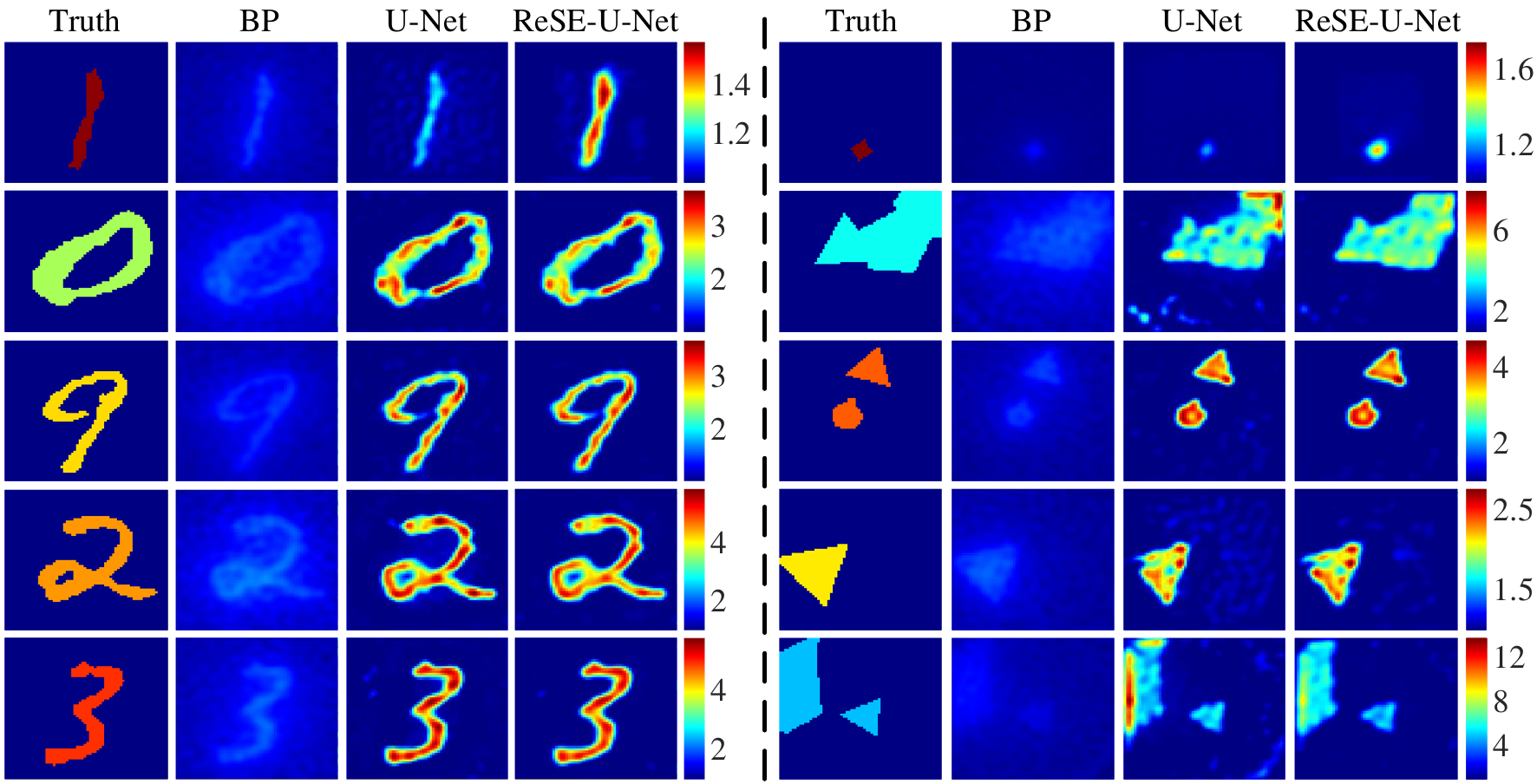}
		\end{center}
		\caption{Image reconstruction of digit-like (left part) and polygon-like (right part) profiles using U-Net and ReSE-U-Net architecture when measured scattered fields are corrupted by Gaussian noises with SNR = 5dB.}
		\label{fig:TestReSE-U-Net}
	\end{figure*}
	
	\begin{figure*}[!th]
		\includegraphics[width=\linewidth]{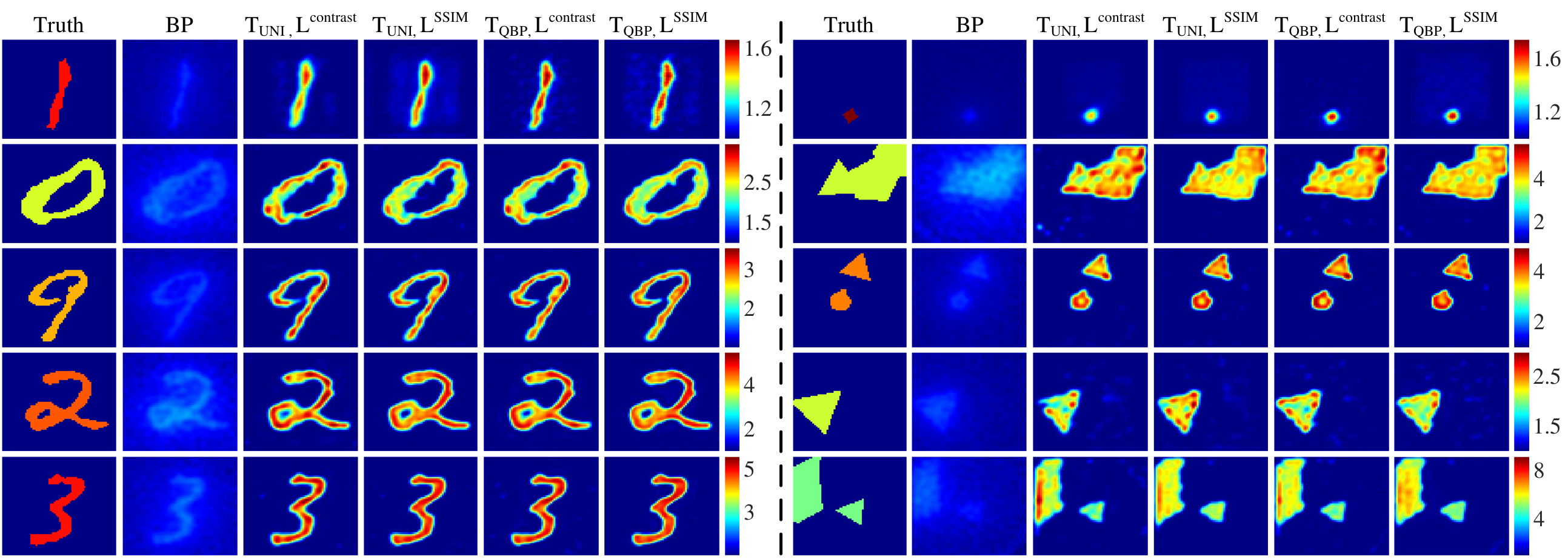}
		\caption{Image reconstruction of digit-like (left part) and polygon-like (right part) scatterers when scattered fields are corrupted by Gaussian noises with SNR = 5dB. The training of ReSE-U-Net was conducted with four different settings: training set $T_{\text{UNI}}$ with loss function $L^{\text{contrast}}$, training set $T_{\text{UNI}}$ with loss function $L^{\text{SSIM}}$, training set $T_{\text{QBP}}$ with loss function $L^{\text{contrast}}$, and training set $T_{\text{QBP}}$ with loss function $L^{\text{SSIM}}$.}
		\label{fig:qualityfactor}
	\end{figure*}
	
	For digit-like and polygon-like scatterers, DNN solvers based on the traditional U-Net architecture and the the ReSE-U-Net one are trained with the training dataset $T_{\text{UNI}}$ when the data are contaminated by white Gaussian noise with SNR = 5dB. The time cost for training U-Net model is 226s and 317s for the ReSE-U-Net model. 2000 digit-like samples and the same number of polygon-like ones are predicted to evaluate the imaging performance.
	
	Five representative digit-like and polygon-like scatterers are shown in Fig.~\ref{fig:TestReSE-U-Net}. As seen, taking the BP result as input, the DNN solvers can significantly improve the imaging performance. For digit-like scatterers, both DNN solvers can accurately reconstruct the geometric feature, but the ReSE-U-Net solver shows superiority in the reconstruction accuracy, especially when observing the imaging results of ``1". Moreover, overestimation tends to happen for high-contrast scatterers. Similar conclusions can be made for polygon-like scatterers, except that the weak and small scatter is more accurately reconstructed with ReSE-U-Net architecture. 
	
	Statistical testing is performed based on 2000 digit-like and polygon-like testing sample and the obtained metrics are presented in Table~\ref{tab:TestReSE-U-Net}. The superiority of applying ReSE-U-Net architecture is proven by observing the mean value of RMSE and SSIM and such superiority is valid for both digit-like and polygon-like scatterers. 
	\begin{table}[!t]
		\centering
		\caption{Statistical Parameters of the Testing Results when using different network architecture.}
		\label{tab:TestReSE-U-Net}
		\begin{tabular}{l|c c|c c}
			\toprule
			\textbf{} & \multicolumn{2}{c|}{\textbf{U-Net}} & \multicolumn{2}{c}{\textbf{ReSE-U-Net}} \\ 
			\midrule
			\textbf{Testing dataset} & \textbf{RMSE} & \textbf{SSIM} & \textbf{RMSE} & \textbf{SSIM} \\ 
			\midrule
			Digit-like   &0.2530 & 0.5692 & 0.2431 &0.7876       \\ 
			Polygon-like &0.2439 & 0.5549 & 0.2201 &0.7718      \\ 
			\bottomrule
		\end{tabular}
	\end{table}
	
	\subsection{Influence from training dataset composition}
	\label{subsec:TestQ}
	Applying the quality factor proposed in Section \ref{sec:DNN solvers} to optimize the composition of training dataset, the five imaging examples same as the ones in Fig.~\ref{fig:TestReSE-U-Net} are given in Fig.~\ref{fig:qualityfactor}. Slight improvement are observed with $T_{\text{QBP}}$ dataset compared with the DNN solver with $T_{\text{UNI}}$ following the observations that details are more accurately reconstructed with $T_{\text{QBP}}$ when imaging number "1" and polygon-like scatterers. The effects from the loss function is more obvious when comparing the imaging results of number “0” and the triangle scatterers where the homogeneous property is more accurately recovered. One also can conclude that the DNN solver with the loss function $L^\text{SSIM}$ trained from the $T_{\text{QBP}}$ is the favorite settings compared with the others.
	
	\begin{table}[!ht]
		\centering
		\caption{Statistical Parameters of Testing Results when using different training datasets.}
		\label{tab:qualityfactor}
		\begin{tabular}{c c|c c|c c}
			\toprule
			\multicolumn{2}{c|}{\textbf{Dataset}} & \multicolumn{2}{c|}{$L^{\text{contrast}}$} & \multicolumn{2}{c}{$L^{\text{SSIM}}$}  \\ 
			\midrule
			\textbf{Train} & \textbf{Test} & \textbf{RMSE} & \textbf{SSIM} & \textbf{RMSE} & \textbf{SSIM}  \\ 
			\midrule
			\multirow{2}{*}{$T_{\text{UNI}}$} & Digit-like   &0.2431 &0.7876 &0.2313 &0.8080 \\ 
			& Polygon-like &0.2201 &0.7718 &0.2126 &0.7803 \\ 
			\midrule
			\multirow{2}{*}{$T_{\text{QBP}}$}  & Digit-like   &0.2314 &0.8073 &0.2253 &0.8436\\ 
			& Polygon-like &0.2165 &0.8074 &0.2121 &0.8086\\ 
			\bottomrule
		\end{tabular}
	\end{table}
	
	Statistical studies are performed based on 2000 digit-like and polygon-like scatterers and the results are given in Table~\ref{tab:qualityfactor}. No matter in terms of RMSE or SSIM, the solver trained with $T_{\text{QBP}}$ dataset has better imaging performances than that with $T_{\text{UNI}}$ dataset for both digit-like testing samples and the polygon-like ones when the same loss function is applied. The SSIM value is increased from 0.8080 to 0.8436 for digit-like scatterers and from 0.7803 to 0.8086 for polygon-like ones after optimizing the composition of training dataset according to the quality factor. Focusing on the effects from loss functions, improvements are obtained that slight decrement are observed and SSIM has an increment of 0.04 when the constraint term about SSIM is in force during the training. The mean value of RMSE and SSIM again reveals the optimal setting of the DNN solver, \emph{i.e.}, the loss function $L^\text{SSIM}$ and the $T_{\text{QBP}}$ dataset.
	Thus, the DNN solver with the optimal setting is further analyzed below.
	
	\subsection{Influence from hyperparameter in loss function}
	\label{subsec:Testhyperparameter }
	
	\begin{figure*}[!th]
		\begin{center}
			\includegraphics[width=0.9\linewidth]{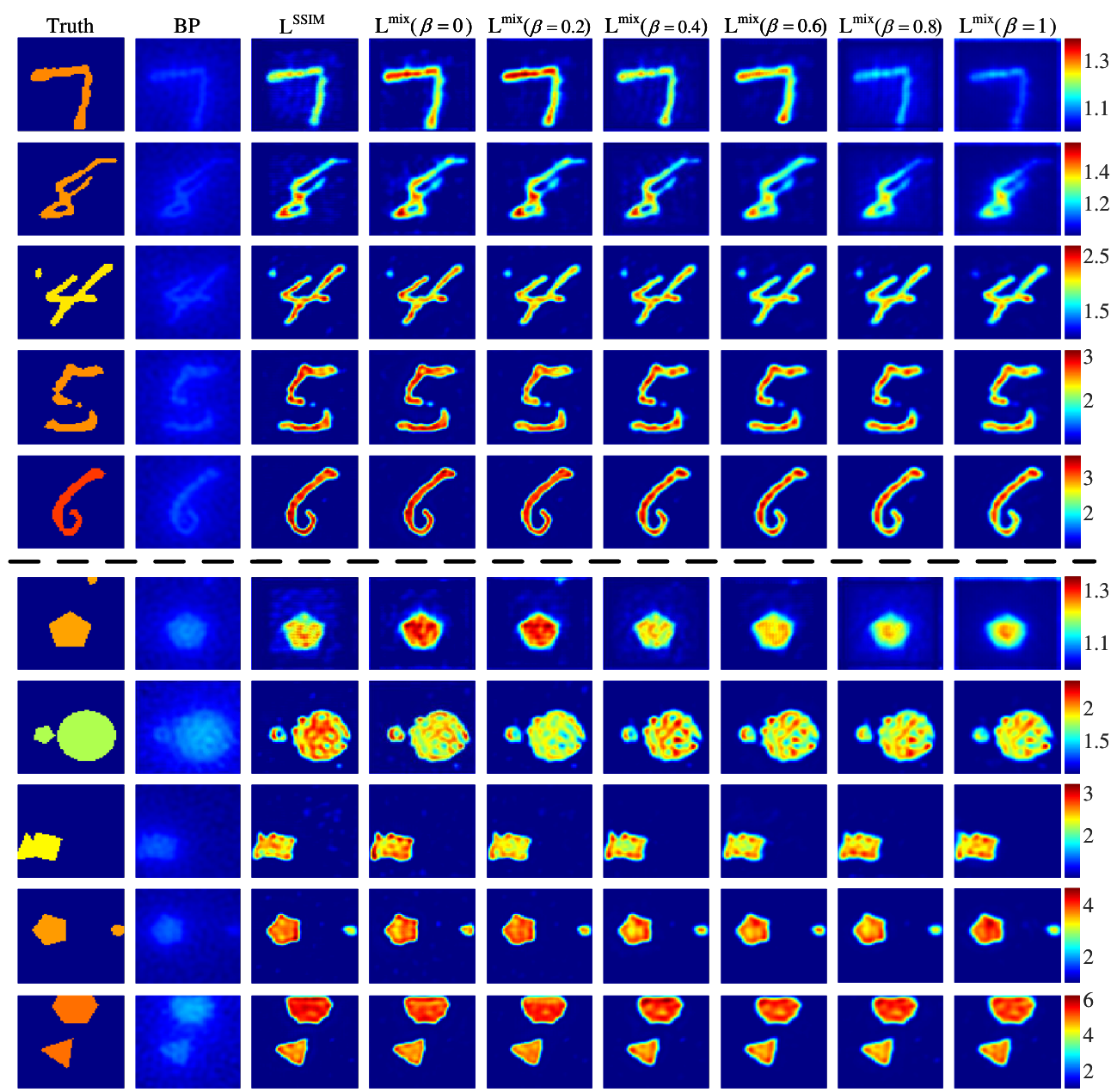}
		\end{center}
		\caption{Imaging results of digit-like scatterers when measured scattered fields are corrupted by Gaussian noises with SNR = 5dB. The training of ReSE-U-Net is based on the training dataset $T_{\text{QBP}}$ with different values of the hyperparameter $\beta$ in the loss function $L^{\text{mix}}$.}
		\label{fig:betaDigit-like}
	\end{figure*}
	
	Two hyperparameters appear in the loss function of (11). While $\alpha$ is determined by an explicit expression, the value of $\beta$ is discussed here. Through numerical analysis, one tends to have an approximation of optimal value of $\beta$. RMSE and SSIM are used to evaluate imaging performances with different values of $\beta$. The results are summarized in Table~\ref{tab:betaDigit-like}. 
	
	\begin{table}[!ht]
		\centering
		\resizebox{1.0\columnwidth}{!}{
		\caption{Statistical Parameters Corresponding to Different Values of Hyperparameter $\beta$ in Loss Function $L^{\text{mix}}$.}
		\label{tab:betaDigit-like}
		\begin{tabular}{c c|c c c c c c c c c c c}
			\toprule
			\textbf{Parameter} & \textbf{testing dataset} & \textbf{$\beta=0$} & \textbf{$\beta=0.1$} & \textbf{$\beta=0.2$} & \textbf{$\beta=0.3$} & \textbf{$\beta=0.4$} & \textbf{$\beta=0.5$} & \textbf{$\beta=0.6$} & \textbf{$\beta=0.7$} & \textbf{$\beta=0.8$} & \textbf{$\beta=0.9$} & \textbf{$\beta=1.0$}  \\ 
			\midrule
			\multirow{2}{*}{RMSE} & Digit-like    &0.2209 &0.2219 &0.2197 &0.2268 &0.2378 &0.2347 &0.2450 &0.2435 &0.2508 &0.2574 &0.2574 \\ 
			& Polygon-like &0.2029 &0.1951 &0.1960 &0.1958 &0.1999 &0.1913 &0.2086 &0.1979 &0.1986 &0.2077 &0.2062 \\ 
			\midrule
			\multirow{2}{*}{SSIM}  & Digit-like   &0.8853 &0.9127 &0.9214 &0.9212 &0.9140 &0.9151 &0.9077 &0.9104 &0.9016 &0.9150 &0.9040 \\ 
			& Polygon-like &0.8665 &0.8932 &0.9026 &0.9048 &0.9137 &0.9022 &0.9095 &0.9076 &0.9114 &0.9281 &0.9177 \\ 
			\bottomrule
		\end{tabular}
		}
	\end{table}
	For digit-like scatterers, with an increase of $\beta$ from 0 to 1, RMSE initially decreases and then gradually increases, while SSIM first increases and then goes into a steady state. These statistical results indicate that imaging performance improves initially but then deteriorates as $\beta$ increases. When $\beta=0$, the loss function $L^{\text{mix}}$ incorporates the physical constraint term $L^{\text{field}}$ and the data-fitting term $L^{\text{SSIM}}$. Comparing the results with $L^{\text{SSIM}}$ along as given in Table~\ref{tab:qualityfactor}, RMSE is slightly reduced while SSIM is increased from 0.8436 to 0.8853. The optimal value of $\beta$ is around 0.2. The statistical results for polygon-like scatterers revealed similar conclusion. Note that the observations that smaller RMSE with polygon-like scatterers than those with digit-like scatterers maybe due to simpler structural features of polygons. The observations also suggest the generalization ability of the obtained DNN solvers. 
	
	The above conclusions can be visually reconfirmed from the ten representative examples in Fig.~\ref{fig:betaDigit-like}. Incorporating the near scattered fields prior knowledge when $\beta=0$ has been improving the reconstruction accuracy, especially when observing the imaging results of digit ``7" and the polygon-like scatterers. The improvement becomes more significant when increasing $\beta$ to 0.2 with the reconstructed relative-permittivity value closer to the ground truth and the homogeneous property of the scatterers is recovered in a higher quality. However, the even larger $\beta$ not necessarily lead to further improvement. When $\beta=1$, the scatter image is overly smoothed. As a result, the image of weak scatterers becomes blurred (e.g., digit ``7" and ``8") and small scatterers are wrongly suppressed in the image of digit ``4".
	
\begin{figure}[!t]
		\begin{center}
			\includegraphics[width= 0.7\linewidth]{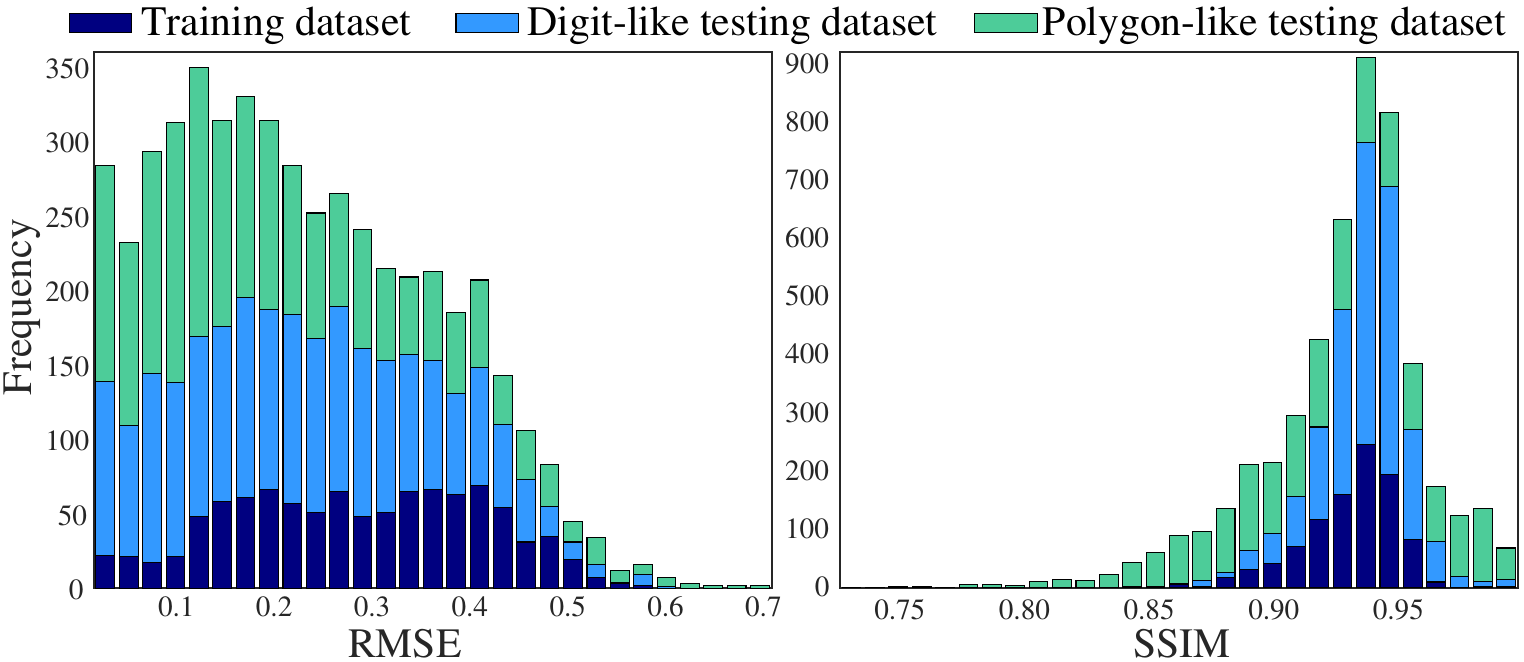}
		\end{center}
		\caption{Distribution of RMSE and SSIM values when predicting samples in the training dataset, the digital-like testing dataset and the polygon-like testing dataset. The loss function $L^{\text{mix}}$ is applied with $\beta=0.2$.}
		\label{fig:histogram}
	\end{figure}

	The scheme of obtaining the DNN solver with the ReSE-U-Net architecture, based on the training dataset $T_{\text{QBP}}$ and applying the the loss function $L^{\text{mix}}$ with the optimal hyperparameter $\beta=0.2$ is QuaDNN($L^{\text{mix}}$). The histogram of QuaDNN solver prediction results is plotted in Fig.~\ref{fig:histogram}. Predicting the samples in the training dataset, the RMSE values are mainly in range of [0.15, 0.4], indicating the reconstruction accuracy. The SSIM values are mainly between 0.90 and 0.95, suggesting high structural similarity between the reconstructed images and the ground truth. For the digit-like testing dataset, the RMSE and SSIM exhibit similar distributions with the training dataset. For the polygon-like testing dataset, most RMSE values are smaller than 0.4 and the SSIM values tend to be uniformly distributed in the range between 0.85 and 1, suggesting the excellent generalization ability of the obtained DNN solver.
	
	\begin{figure*}[!th]
		\begin{center}
			\includegraphics[width=\linewidth]{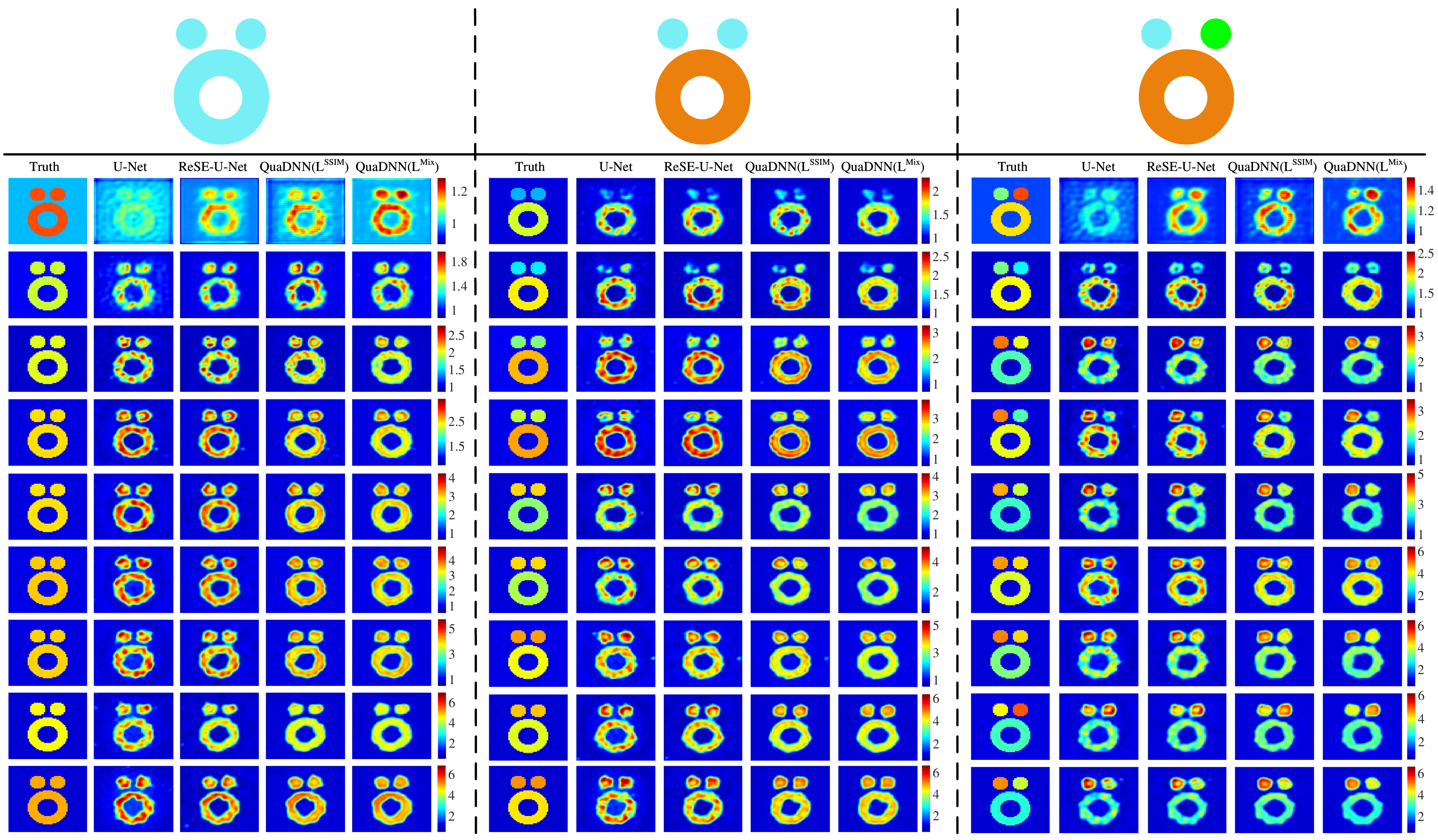}
		\end{center}
		\caption{Imaging results of Austria profiles with different values of relative permittivity. The training dataset and the testing dataset are both with SNR = 5dB.}
		\label{fig:Austria}
	\end{figure*}
	
	\subsection{Testing with Austria profile}
	\label{subsec:Austria}
	The Austria profile consists of two circles with the same radius of $0.56\lambda_0$ and centered at $[\pm0.7\lambda_0, 1.4\lambda_0]$, and a annulus characterized with inner radius of $0.7\lambda_0$ and outer radius of $1.4\lambda_0$. As shown in Fig.~\ref{fig:Austria}, the relative permittivity distribution is varied in three ways, varying the entire profile simultaneously, varying the two circles simultaneously, and varying the three components independently. Fig.~\ref{fig:Austria} presents the reconstruction results when SNR = 5dB. 
	
	When the whole profile is homogeneous and with low contrast, the U-Net model failed to reconstruct the profile, while ReSE-U-Net and QuaDNN($L^\text{SSIM}$) underestimate the contrast values. In contrast, QuaDNN($L^\text{Mix}$) achieves accurate reconstruction and performs the best among all schemes. As the relative permittivity increases, although the nonlinearity extent is higher, the reconstruction performance of the QuaDNN($L^\text{Mix}$) can always recover the shape and the contrast value accurately and outperforms the other solvers. Such phenomenon proves the wide applicability and stability of the proposed QuaDNN($L^\text{Mix}$) training scheme. 
	
	Varying the relative permittivity of the two circles, the whole profile becomes inhomogeneous and the cases are valuable for the generalization ability testing considering that only homogeneous samples are considering in the training dataset. When the contrast of the circles is low, since the contribution from the circles on the scattered fields is small compared with that from the annulus, the shape of the circles are not fully recovered. The solvers with U-Net and ReSE-U-Net overestimate the contrast of the annulus.
	
	When the two circles and the annulus are independent for the contrast variation, one observe that under low SNR conditions, it is still challenging to reconstruct lowest contrast scatterers. With U-Net and ReSE-U-Net, the image of the three components sometimes are overlapped, but such overlapping is successfully suppressed with QuaDNN($L^\text{Mix}$) solver.
	
	\subsection{Testing with overlapping profiles}
	\label{subsec:overlap}
\begin{figure}[!th]
		\begin{center}
			\includegraphics[width=0.6\linewidth]{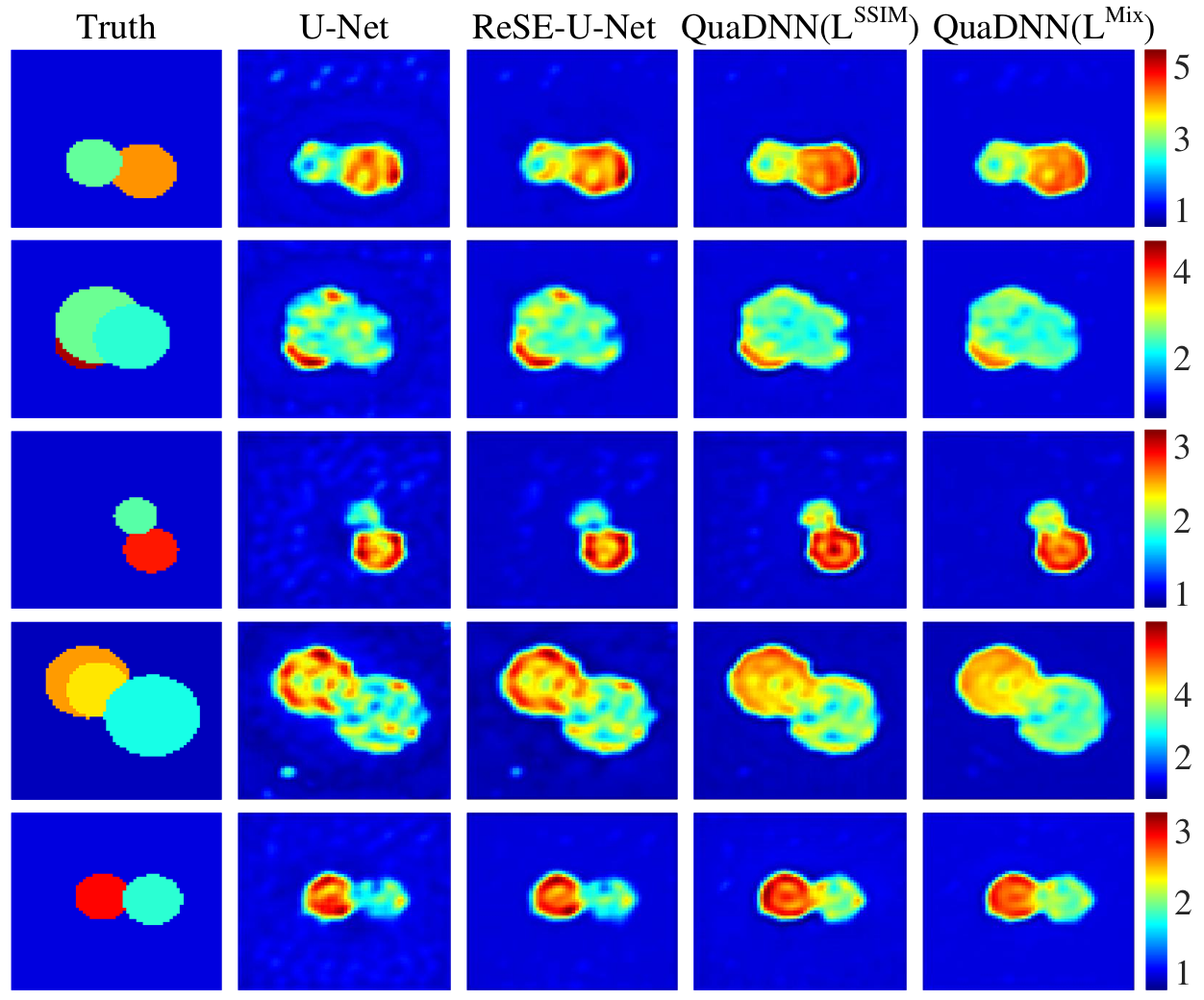}
		\end{center}
		\caption{Reconstructed overlapping circular profiles. The training dataset and the testing dataset are both with SNR = 5dB.}
		\label{fig:overlap}
	\end{figure}

	To further evaluate the robustness of the DNN solvers under strong nonlinearity conditions, overlapping scatterers with different relative permittivity values are designed and used for testing. The results are shown in Fig.~\ref{fig:overlap}. The U-Net and ReSE-U-Net solvers fail to reconstruct the piece-wise homogeneous property and background artifacts appear. With QuaDNN($L^{\text{SSIM}}$), the background artifacts are suppressed and the edge of scatterers are more precisely reconstructed. The imaging performance gets improved even more with QuaDNN($L^\text{Mix}$), where the homogeneous property is closer to the ground truth and overlapping regions with distinct permittivity are well-resolved. 
	
	\subsection{SNR mismatching effects}
	\label{subsec:mismatch}
	\begin{figure}[!th]
		\begin{center}
			\includegraphics[width=\linewidth]{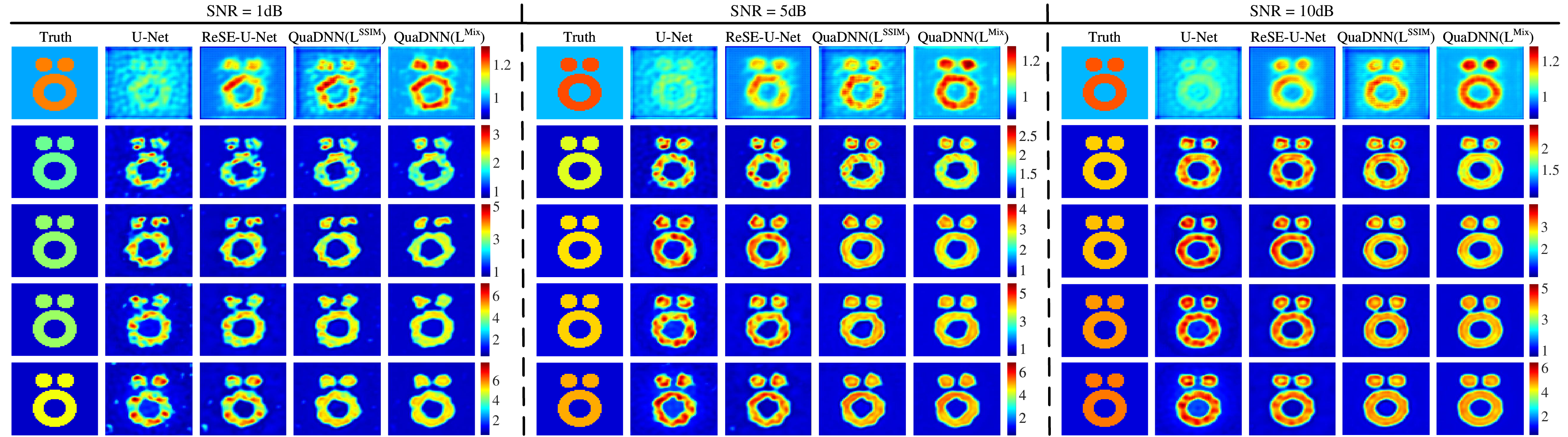}
		\end{center}
		\caption{Study of SNR mismatching effects 
			by varying the SNR value of the testing dataset of Austria profile as 1dB, 5dB and 10dB, while the training dataset is with SNR = 5dB.}
		\label{fig:mismatchSNR}
	\end{figure}
	
	Fig.~\ref{fig:mismatchSNR} presents the reconstruction results for testing datasets with different SNR values, while the DNN solvers are obtained based on training dataset with SNR = 5dB. When SNR = 1dB, the imaging performance deteriorates for all DNN solvers. With U-Net, the scatter image is submerged by background artifacts for the low-contrast case and the reconstructed contrast profile has significant distinction with the ground truth, despite the geometry is well recovered. The imaging performance improves with ReSE-U-Net and QuaDNN($L^{\text{SSIM}}$), but the QuaDNN($L^\text{Mix}$) again proved its superiority with advantages in the reconstruction accuracy and applicability. As the SNR increases to 10dB, all solvers benefit from the reduced noise level, leading to better reconstruction quality. However, for low-contrast scatterers, the U-Net still performs poorly, while ReSE-U-Net produces clearer boundaries of the scatterers. QuaDNN($L^\text{Mix}$) performs the best and can well reconstruct both weak and strong scatterers. In summary, QuaDNN($L^\text{Mix}$) scheme shows the robustness to SNR mismatching.
	
	\subsection{Testing with experimental data}
	\label{subsec:experimental}
	
	\begin{figure}[!t]
		\begin{center}
			\includegraphics[width=0.3\linewidth]{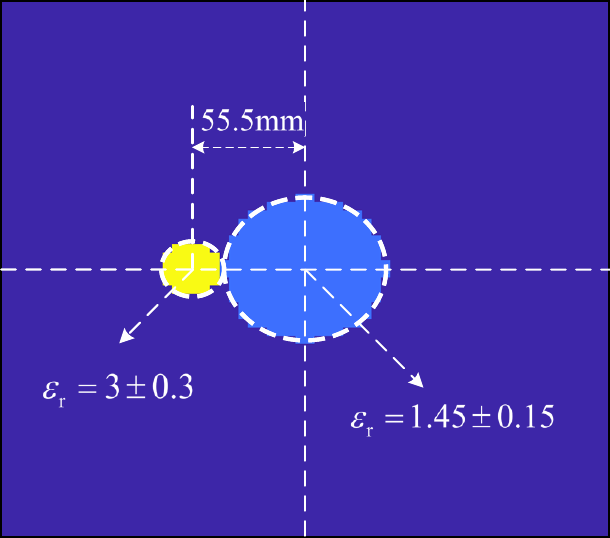}
		\end{center}
		\caption{Ground truth of scatter image for “FoamDielExt”experimental data. }
		\label{fig:ExpGroundTruth}
	\end{figure}
	
	\begin{figure}[!t]
		\begin{center}
			\includegraphics[width=0.5\linewidth]{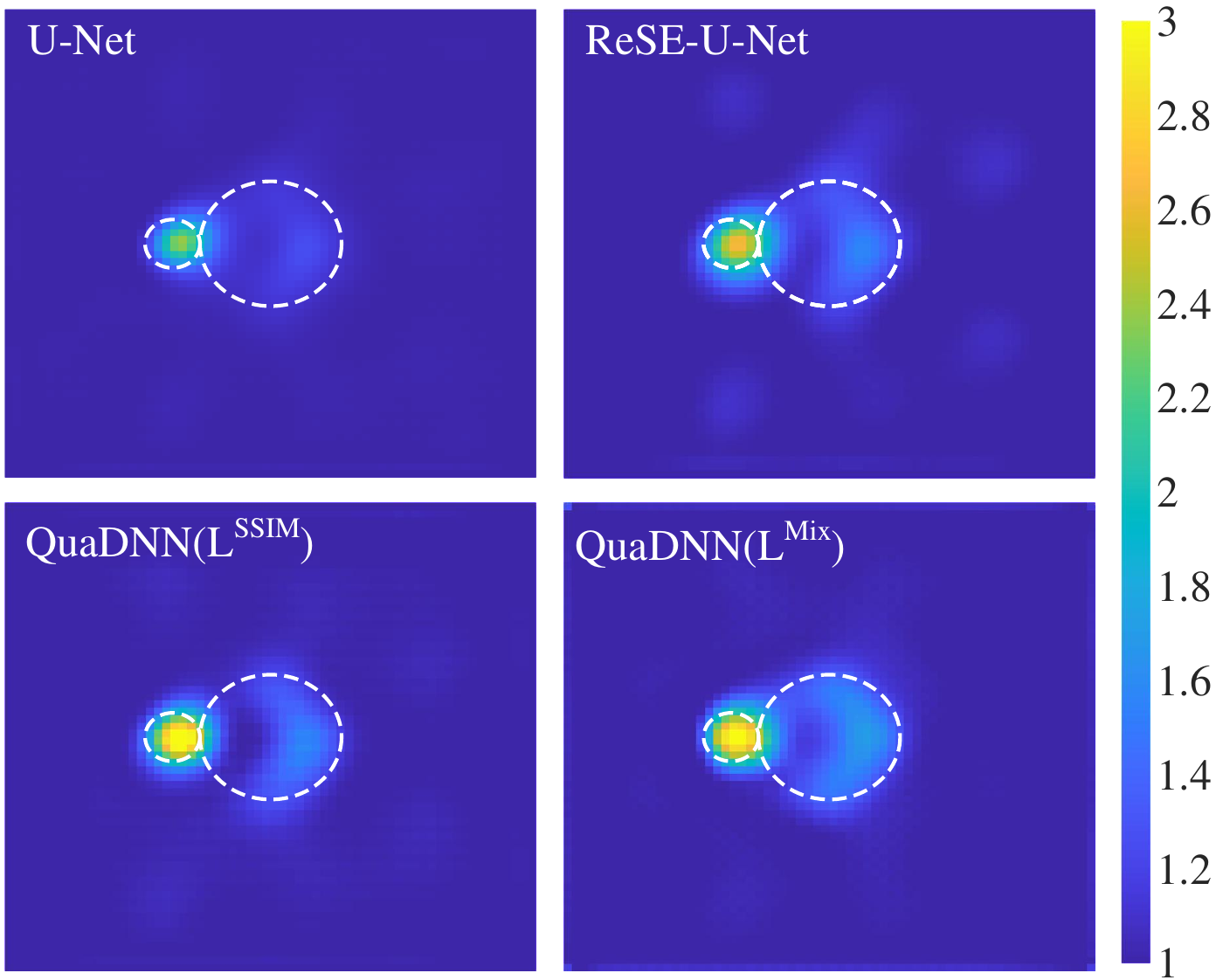}
		\end{center}
		\caption{Imaging the two cylinders from the experimental data “FoamDielExt”.  }
		\label{fig:ExpPredict}
	\end{figure}
	
	The imaging performance of the obtained DNN solvers are verified with the experimental data provided by Fresnel Institute in Marseille, France \cite{geffrin2005Fresnel}. As shown in Fig.~\ref{fig:ExpGroundTruth}, The experiment includes two cylindrical scatterers, the radius of the left cylinder is 15.5 mm and the right one 40 mm. The relative permittivity of the left cylinder is 1.45 ± 0.15, and the right one is 3 ± 0.3. The symbol “±” denotes the permissible range of measurement error or uncertainty in the values. The distance between the centers of the two cylinders is 55.5 mm. 
	
	The reconstructed results of the “FoamDielExt” scatterers are shown in Fig.~\ref{fig:ExpPredict}. The four images respectively show the reconstruction results of solvers with U-Net, ReSE-U-Net, QuaDNN($L^{\text{SSIM}}$) and QuaDNN($L^\text{Mix}$). With U-Net, the relative permittivity of two cylinders is underestimated and the edge of the right cylinder is blurred. With ReSE-U-Net, the high-contrast cylinder is better imaged with improved reconstruction accuracy. With QuaDNN($L^{\text{SSIM}}$), the size and the permittivity value of the left cylinder is accurately reconstructed, but the low-contrast one is not fully recovered. With QuaDNN($L^\text{Mix}$), despite the perfect imaging of the left cylinder, more details are recovered about the right one, although the cylinder shape is still not fully imaged.
	
	\section{Conclusions}
	\label{sec:conclusions}
	In this paper, a training scheme for the DNN ISP solver is proposed. Integrating the residual, SE blocks and feature transformation layer into a U-Net neural network to have the so-call ReSE-U-Net architecture, optimizing the training dataset based on the defined quality factor, and applying the loss function which combines the data-fitting term, near-scattered-field constraints and the smoothness prior information of the desired solution, superior imaging performances are achieved and proven via comparisons with the traditional method BP and the DNN solvers trained based on different schemes. 
	
	The imaging performances are demonstrated by testing the polygon-like scatterers, Austria profiles, and overlapping circle scatterers to validate the generalization ability. Enhanced geometric features and more accurate relative permittivity reconstruction are observed. The robustness to different noise levels and the SNR mismatching is also tested through numerical analysis. Such observations imply the effectiveness of the proposed training scheme.
	
	In this paper, the proportion of samples from different quality sets was fixed, in order to demonstrate the benefits when optimizing the composition of training dataset. However, the optimal setting for the proportion is possible and will be studied in the future.
	
	\bibliographystyle{chicago}
	\bibliography{DNN}

\begin{thebibliography}{}

\bibitem[\protect\citeauthoryear{An, Li, and Ding}{An et~al.}{2024}]{An2024Composites}
An, K., C.~Li, and J.~Ding (2024, August).
\newblock Non-destructive testing of multi-layer composites based on microwave time reversal.
\newblock In {\em Proc. Int. Appl. Comput. Electromagn. Soc. Symp.}

\bibitem[\protect\citeauthoryear{Belkebir, Chaumet, and Sentenac}{Belkebir et~al.}{2005}]{Belkebir2005BP}
Belkebir, K., P.~C. Chaumet, and A.~Sentenac (2005, Sep).
\newblock Superresolution in total internal reflection tomography.
\newblock {\em J. Opt. Soc. Am. A\/}~{\em 22\/}(9), 1889--1897.

\bibitem[\protect\citeauthoryear{Bloemenkamp, Abubakar, and van~den Berg}{Bloemenkamp et~al.}{2001}]{richard2001CSI}
Bloemenkamp, R.~F., A.~Abubakar, and P.~M. van~den Berg (2001).
\newblock Inversion of experimental multi-frequency data using the contrast source inversion method.
\newblock {\em Inverse Probl.\/}~{\em 17\/}(6), 1611.

\bibitem[\protect\citeauthoryear{Bucci and Isernia}{Bucci and Isernia}{1997}]{bucci1997quantitativeanalysis}
Bucci, O.~M. and T.~Isernia (1997).
\newblock Electromagnetic inverse scattering: Retrievable information and measurement strategies.
\newblock {\em Radio Sci.\/}~{\em 32\/}(6), 2123--2137.

\bibitem[\protect\citeauthoryear{Chen}{Chen}{2010}]{chen2010SOM}
Chen, X. (2010).
\newblock Subspace-based optimization method for solving inverse-scattering problems.
\newblock {\em IEEE Trans. Geosci. Remote Sens.\/}~{\em 48\/}(1), 42--49.

\bibitem[\protect\citeauthoryear{Chen}{Chen}{2018}]{chen2018computationalEMIS}
Chen, X. (2018).
\newblock {\em Computational Methods for Electromagnetic Inverse Scattering}.
\newblock Hoboken, NJ, USA: Wiley.

\bibitem[\protect\citeauthoryear{Chen, Zhang, Cui, Teixeira, and Li}{Chen et~al.}{2023}]{Chen2023CNN}
Chen, Y., H.~Zhang, T.~J. Cui, F.~L. Teixeira, and L.~Li (2023).
\newblock A mesh-free 3-{D} deep learning electromagnetic inversion method based on point clouds.
\newblock {\em IEEE Trans. Microw. Theory Tech.\/}~{\em 71\/}(8), 3530--3539.

\bibitem[\protect\citeauthoryear{Chew and Wang}{Chew and Wang}{1990}]{chew1990DBIM}
Chew, W. and Y.~Wang (1990).
\newblock Reconstruction of two-dimensional permittivity distribution using the distorted born iterative method.
\newblock {\em IEEE Trans. Med. Imaging\/}~{\em 9\/}(2), 218--225.

\bibitem[\protect\citeauthoryear{Dachena, Fedeli, Fanti, Lodi, Fumera, Randazzo, and Pastorino}{Dachena et~al.}{2021}]{dachena2021tumordetection}
Dachena, C., A.~Fedeli, A.~Fanti, M.~B. Lodi, G.~Fumera, A.~Randazzo, and M.~Pastorino (2021).
\newblock Microwave imaging of the neck by means of artificial neural networks for tumor detection.
\newblock {\em IEEE Open J. Antennas Propag.\/}~{\em 2}, 1044--1056.

\bibitem[\protect\citeauthoryear{Devaney}{Devaney}{1981}]{devaney1981RA}
Devaney, A.~J. (1981).
\newblock Inverse-scattering theory within the rytov approximation.
\newblock {\em Opt. Lett.\/}~{\em 6\/}(8), 374--376.

\bibitem[\protect\citeauthoryear{Devaney}{Devaney}{1982}]{devaney1982BP}
Devaney, A.~J. (1982).
\newblock A filtered backpropagation algorithm for diffraction tomography.
\newblock {\em Ultrason. Imaging\/}~{\em 4\/}(4), 336--350.

\bibitem[\protect\citeauthoryear{Gao and Torres-Verdin}{Gao and Torres-Verdin}{2006}]{gao2006BA}
Gao, G. and C.~Torres-Verdin (2006).
\newblock High-order generalized extended born approximation for electromagnetic scattering.
\newblock {\em IEEE Trans. Antennas Propag.\/}~{\em 54\/}(4), 1243--1256.

\bibitem[\protect\citeauthoryear{Geffrin, Sabouroux, and Eyraud}{Geffrin et~al.}{2005}]{geffrin2005Fresnel}
Geffrin, J.-M., P.~Sabouroux, and C.~Eyraud (2005).
\newblock Free space experimental scattering database continuation: experimental set-up and measurement precision.
\newblock {\em Inverse Probl.\/}~{\em 21\/}(6), S117.

\bibitem[\protect\citeauthoryear{Gibson}{Gibson}{2021}]{Gibson2021MoM}
Gibson, W. (2021).
\newblock {\em The Method of Moments in Electromagnetics}.
\newblock New York, NY, USA: Chapman and Hall/CRC.

\bibitem[\protect\citeauthoryear{Golub, Hansen, and O'Leary}{Golub et~al.}{1999}]{golub1999Tikhonov}
Golub, G.~H., P.~C. Hansen, and D.~P. O'Leary (1999).
\newblock Tikhonov regularization and total least squares.
\newblock {\em SIAM J. Matrix Anal. Appl.\/}~{\em 21\/}(1), 185--194.

\bibitem[\protect\citeauthoryear{Habashy, Groom, and Spies}{Habashy et~al.}{1993}]{habashyi1993BA}
Habashy, T.~M., R.~W. Groom, and B.~R. Spies (1993).
\newblock Beyond the born and rytov approximations: A nonlinear approach to electromagnetic scattering.
\newblock {\em J. Geophys. Res. Solid Earth\/}~{\em 98\/}(B2), 1759--1775.

\bibitem[\protect\citeauthoryear{Haddadin, Lucas, and Ebbini}{Haddadin et~al.}{1995}]{haddadin1995DBIM}
Haddadin, O., S.~Lucas, and E.~Ebbini (1995).
\newblock Solution to the inverse scattering problem using a modified distorted born iterative algorithm.
\newblock In {\em 1995 IEEE Ultrason. Symp. Proc.}, Volume~2, pp.\  1411--1414.

\bibitem[\protect\citeauthoryear{He, Zhang, Ren, and Sun}{He et~al.}{2016}]{He2016residual}
He, K., X.~Zhang, S.~Ren, and J.~Sun (2016, June).
\newblock Deep residual learning for image recognition.
\newblock In {\em Proc. IEEE Conf. Comput. Vis. Pattern Recognit.}

\bibitem[\protect\citeauthoryear{Hirose, Zhu, and Kidera}{Hirose et~al.}{2022}]{hirose2022cancerdiagnosis}
Hirose, U., P.~Zhu, and S.~Kidera (2022).
\newblock Deep learning enhanced contrast source inversion for microwave breast cancer imaging modality.
\newblock {\em IEEE J. Electromagn. RF Microw. Med. Biol.\/}~{\em 6\/}(3), 373--379.

\bibitem[\protect\citeauthoryear{Hu, Shen, and Sun}{Hu et~al.}{2018}]{Hu2018SE}
Hu, J., L.~Shen, and G.~Sun (2018, June).
\newblock Squeeze-and-excitation networks.
\newblock In {\em Proc. IEEE Conf. Comput. Vis. Pattern Recognit.}

\bibitem[\protect\citeauthoryear{Jun and Choi}{Jun and Choi}{1999}]{sung1999BIM}
Jun, S.~C. and U.~J. Choi (1999).
\newblock Convergence analyses of the born iterative method and the distorted born iterative method.
\newblock {\em Numer. Funct. Anal. Optim.\/}~{\em 20\/}(3-4), 301--316.

\bibitem[\protect\citeauthoryear{Lakhtakia}{Lakhtakia}{1992}]{LAKHTAKIA1992MoM}
Lakhtakia, A. (1992).
\newblock Strong and weak forms of the method of moments and the coupled dipole method for scattering of time-harmonic electromagnetic fields.
\newblock {\em Int. J. Mod. Phys. C\/}~{\em 3\/}(2), 583--603.

\bibitem[\protect\citeauthoryear{Lecun, Bottou, Bengio, and Haffner}{Lecun et~al.}{1998}]{Lecun1998MNIST}
Lecun, Y., L.~Bottou, Y.~Bengio, and P.~Haffner (1998).
\newblock Gradient-based learning applied to document recognition.
\newblock {\em Proc. IEEE\/}~{\em 86\/}(11), 2278--2324.

\bibitem[\protect\citeauthoryear{Li, Wang, Teixeira, Liu, Nehorai, and Cui}{Li et~al.}{2019}]{li2019DNN}
Li, L., L.~G. Wang, F.~L. Teixeira, C.~Liu, A.~Nehorai, and T.~J. Cui (2019).
\newblock Deep{NIS}: Deep neural network for nonlinear electromagnetic inverse scattering.
\newblock {\em IEEE Trans. Antennas Propag.\/}~{\em 67\/}(3), 1819--1825.

\bibitem[\protect\citeauthoryear{Liu, Roy, Prasad, and Agarwal}{Liu et~al.}{2022}]{liu2022DNN}
Liu, Z., M.~Roy, D.~K. Prasad, and K.~Agarwal (2022).
\newblock Physics-guided loss functions improve deep learning performance in inverse scattering.
\newblock {\em IEEE Trans. Comput. Imaging\/}~{\em 8}, 236--245.

\bibitem[\protect\citeauthoryear{Ma, Liu, and Zong}{Ma et~al.}{2023}]{ma2023DNN}
Ma, J., Z.~Liu, and Y.~Zong (2023).
\newblock Inverse scattering solver based on deep neural network with total variation regularization.
\newblock {\em IEEE Antennas Wireless Propag. Lett.\/}~{\em 22\/}(10), 2447--2451.

\bibitem[\protect\citeauthoryear{Meaney, Paulsen, and Ryan}{Meaney et~al.}{1995}]{meaney1995TM}
Meaney, P., K.~Paulsen, and T.~Ryan (1995).
\newblock Two-dimensional hybrid element image reconstruction for {TM} illumination.
\newblock {\em IEEE Trans. Antennas Propag.\/}~{\em 43\/}(3), 239--247.

\bibitem[\protect\citeauthoryear{Nachman}{Nachman}{1988}]{Adrian1988quant}
Nachman, A.~I. (1988).
\newblock Reconstructions from boundary measurements.
\newblock {\em Ann. of Math.\/}~{\em 128\/}(3), 531--576.

\bibitem[\protect\citeauthoryear{Ney}{Ney}{1985}]{Ney1985MoM}
Ney, M. (1985).
\newblock Method of moments as applied to electromagnetic problems.
\newblock {\em IEEE Trans. Microw. Theory Tech.\/}~{\em 33\/}(10), 972--980.

\bibitem[\protect\citeauthoryear{Pan, Zhong, Chen, and Yeo}{Pan et~al.}{2011}]{pan2011SOM}
Pan, L., Y.~Zhong, X.~Chen, and S.~P. Yeo (2011).
\newblock Subspace-based optimization method for inverse scattering problems utilizing phaseless data.
\newblock {\em IEEE Trans. Geosci. Remote Sens.\/}~{\em 49\/}(3), 981--987.

\bibitem[\protect\citeauthoryear{Ramm}{Ramm}{1988}]{A1988quant}
Ramm, A.~G. (1988).
\newblock Recovery of the potential from fixed-energy scattering data.
\newblock {\em Inverse Probl.\/}~{\em 4\/}(3), 877.

\bibitem[\protect\citeauthoryear{Rocca, Benedetti, Donelli, Franceschini, and Massa}{Rocca et~al.}{2009}]{P2009quant}
Rocca, P., M.~Benedetti, M.~Donelli, D.~Franceschini, and A.~Massa (2009).
\newblock Evolutionary optimization as applied to inverse scattering problems.
\newblock {\em Inverse Probl.\/}~{\em 25\/}(12), 123003.

\bibitem[\protect\citeauthoryear{Ronneberger, Fischer, and Brox}{Ronneberger et~al.}{2015}]{ronneberger2015UNet}
Ronneberger, O., P.~Fischer, and T.~Brox (2015).
\newblock U-net: Convolutional networks for biomedical image segmentation.
\newblock In {\em Proc. 18th Int. Conf. Med. ImageComput. Comput.-Assist. Intervent.}, pp.\  234--241.

\bibitem[\protect\citeauthoryear{Rudin, Osher, and Fatemi}{Rudin et~al.}{1992}]{Leonid1992TV}
Rudin, L.~I., S.~Osher, and E.~Fatemi (1992).
\newblock Nonlinear total variation based noise removal algorithms.
\newblock {\em Phys. D\/}~{\em 60\/}(1), 259--268.

\bibitem[\protect\citeauthoryear{Schatzberg and Devaney}{Schatzberg and Devaney}{1993}]{alon1993RA}
Schatzberg, A. and A.~J. Devaney (1993).
\newblock Rough surface inverse scattering within the rytov approximation.
\newblock {\em J. Opt. Soc. Am. A\/}~{\em 10\/}(5), 942--950.

\bibitem[\protect\citeauthoryear{Slaney, Kak, and Larsen}{Slaney et~al.}{1984}]{slaney1984RA}
Slaney, M., A.~Kak, and L.~Larsen (1984).
\newblock Limitations of imaging with first-order diffraction tomography.
\newblock {\em IEEE Trans. Microwave Theory Tech.\/}~{\em 32\/}(8), 860--874.

\bibitem[\protect\citeauthoryear{Stefanov}{Stefanov}{1990}]{Stefanov1990quant}
Stefanov, P. (1990).
\newblock Stability of the inverse problem in potential scattering at fixed energy.
\newblock {\em Ann. Inst. Fourier\/}~{\em 40\/}(4), 867--884.

\bibitem[\protect\citeauthoryear{van~den Berg and Kleinman}{van~den Berg and Kleinman}{1997}]{peter1997CSI}
van~den Berg, P.~M. and R.~E. Kleinman (1997).
\newblock A contrast source inversion method.
\newblock {\em Inverse Probl.\/}~{\em 13\/}(6), 1607.

\bibitem[\protect\citeauthoryear{Wang and Oristaglio}{Wang and Oristaglio}{1998}]{tsili1998BP}
Wang, T. and M.~L. Oristaglio (1998).
\newblock An inverse algorithm for velocity reconstruction.
\newblock {\em Inverse Probl.\/}~{\em 14\/}(5), 1345.

\bibitem[\protect\citeauthoryear{Wang and Chew}{Wang and Chew}{1989}]{wang1989BIM}
Wang, Y.~M. and W.~C. Chew (1989).
\newblock An iterative solution of the two-dimensional electromagnetic inverse scattering problem.
\newblock {\em Int. J. Imag. Syst. Technol.\/}~{\em 1\/}(1), 100--108.

\bibitem[\protect\citeauthoryear{Wei and Chen}{Wei and Chen}{2019}]{wei2019DNN}
Wei, Z. and X.~Chen (2019).
\newblock Deep-learning schemes for full-wave nonlinear inverse scattering problems.
\newblock {\em IEEE Trans. Geosci. Remote Sens.\/}~{\em 57\/}(4), 1849--1860.

\bibitem[\protect\citeauthoryear{Wu, Peng, Wang, Wang, and Xiang}{Wu et~al.}{2024}]{Wu2024CNN}
Wu, Z., Y.~Peng, P.~Wang, W.~Wang, and W.~Xiang (2024).
\newblock A physics-induced deep learning scheme for electromagnetic inverse scattering.
\newblock {\em IEEE Trans. Microw. Theory Tech.\/}~{\em 72\/}(2), 927--947.

\bibitem[\protect\citeauthoryear{{X. Chen}}{{X. Chen}}{2010}]{chen_2010SOM}
{X. Chen} (2010).
\newblock Subspace-based optimization method for inverse scattering problems with an inhomogeneous background medium.
\newblock {\em Inverse Probl.\/}~{\em 26\/}(7), 074007.

\bibitem[\protect\citeauthoryear{Yao, Sha, and Jiang}{Yao et~al.}{2019}]{yao2019DNN}
Yao, H.~M., W.~E.~I. Sha, and L.~Jiang (2019).
\newblock Two-step enhanced deep learning approach for electromagnetic inverse scattering problems.
\newblock {\em IEEE Antennas Wireless Propag. Lett.\/}~{\em 18\/}(11), 2254--2258.

\end{thebibliography}
\end{document}